\documentclass[runningheads]{llncs}
\usepackage{graphicx}
\usepackage{multirow}
\usepackage[table,xcdraw]{xcolor}
\usepackage[normalem]{ulem}
\useunder{\uline}{\ul}{} 
\usepackage{multirow}
\usepackage {xcolor}
\usepackage{hyperref}
\usepackage{amsmath}
\usepackage{amssymb}
\usepackage{bm}
\usepackage{cancel}
\usepackage{color, colortbl}
\usepackage{subcaption}
\usepackage[misc]{ifsym}
\usepackage{booktabs} 
\usepackage{xcolor}
\definecolor{Gray}{gray}{0.9}
\definecolor{LightCyan}{rgb}{0.88,1,1}

\begin{document}

\title{PEMMA: Parameter-Efficient Multi-Modal Adaptation for Medical Image Segmentation}

\titlerunning{}
%
\author{Nada Saadi \and Numan Saeed  \and Mohammad Yaqub  \and Karthik Nandakumar }
\authorrunning{N.Saadi et al.}
\institute{Mohamed bin Zayed University of Artificial Intelligence, Abu Dhabi, UAE
\email{\{ nada.saadi, numan.saeed, mohammad.yaqub, karthik.nandakumar\}@mbzuai.ac.ae}
}

\maketitle              
\begin{abstract}

Imaging modalities such as Computed Tomography (CT) and Positron Emission Tomography (PET) are key in cancer detection, inspiring Deep Neural Networks (DNN) models that merge these scans for tumor segmentation. When both CT and PET scans are available, it is common to combine them as two channels of the input to the segmentation model. However, this method requires both scan types during training and inference, posing a challenge due to the limited availability of PET scans, thereby sometimes limiting the process to CT scans only. Hence, there is a need to develop a flexible DNN architecture that can be trained/updated using only CT scans but can effectively utilize PET scans when they become available. In this work, we propose a \underline{\textbf{p}}arameter-\underline{\textbf{e}}fficient \underline{\textbf{m}}ulti-\underline{\textbf{m}}odal \underline{\textbf{a}}daptation (PEMMA) framework for lightweight upgrading of a transformer-based segmentation model trained only on CT scans to also incorporate PET scans. The benefits of the proposed approach are two-fold. Firstly, we leverage the inherent modularity of the transformer architecture and perform low-rank adaptation (LoRA) of the attention weights to achieve \textit{parameter-efficient} adaptation. Secondly, since the PEMMA framework attempts to \textit{minimize cross-modal entanglement}, it is possible to subsequently update the combined model using only one modality, without causing catastrophic forgetting of the other modality. Our proposed method achieves comparable results with the performance of early fusion techniques with just 8\%  of the trainable parameters, especially with a remarkable +28\%  improvement on the average dice score on PET scans when trained on a single modality.


\end{abstract}

{

\section{Introduction}

In clinical practice, precisely identifying cancerous tumors through various imaging techniques presents a significant challenge due to the intricate nature of the disease. The necessity to interpret numerous imaging modalities complicates and lengthens the diagnostic and prognostic process, often making it arduous and monotonous. Consequently, there is a critical need for the advancement of Deep Neural Network (DNN) models that can automate this process. Moreover, there is a need to refine the integration of diverse modalities, mirroring the nuanced fusion approach instinctively employed by oncologists, thereby substantially enhancing accuracy and minimizing the intra- and inter-observer variability. 

It is well-known that integration of both CT and PET imaging modalities using DNNs can significantly enhance the accuracy of tumor segmentation \cite{doi:fusion-10.1080/0284186X.2021.1949034}. This can be attributed to the comprehensive evaluation of anatomical structures by the CT scans and the ability of PET to capture metabolic activity. However, the availability of PET scans may be restricted in practice due to cost and clinical necessity. Joint processing of these two imaging modalities has been extensively studied in the literature \cite{TMSS-10.1007/978-3-031-16449-1_31}, \cite{fusion-huang2024vision}. These approaches involve either merging the modalities at the input level as channels \cite{early-fusion10.1055/a-1542-6211} or at the output level after independent processing of each modality \cite{late-fusion10.1117/12.878067}. While the former approach assumes constant availability of both modalities, the latter approach doubles the number of trained parameters \cite{el-wang2022deep}. Hence, there is a need for more sophisticated method for jointly handling CT and PET scans.

In terms of DNN architecture used to process CT and PET modalities, several models such as Convolutional Neural Networks (CNNs), Graph Neural Networks, and transformers have been studied in the literature \cite{gnn-PENG2024106137}. While CNN-based models have demonstrated impressive performance in various applications, opting for transformer-based models for segmentation tasks offers several advantages. Specifically, due to their unparalleled ability to capture long-range dependencies and intricate patterns within the data, transformer based models can lead to more refined and accurate outcomes. In this context, the UNETR model \cite{hatamizadeh2021unetr} marks a significant departure from traditional CNNs and uses the Vision Transformer (ViT) architecture. In general, the image encoder in a transformer-based segmentation model converts the given image $\mathbf{x}$ into a set of $N$ patch tokens $\mathcal{T}_0 =\{\mathbf{t}_1,\cdots,\mathbf{t}_N\}$ via a patch embedding layer $\mathcal{E}_{\theta_{\textrm{PE}}}$, where $\theta_{\textrm{PE}}$ denotes the patch embedding parameters. These tokens are processed through a sequence of $L$ identical self-attention transformer blocks, where the operation of each block can be expressed as $\mathcal{T}_{\ell} = \mathcal{G}_{\theta_{\ell}}(\mathcal{T}_{\ell-1})$, where $\theta_{\ell}$ represents the parameters of the $\ell^{th}$ transformer block, $\ell  \in [1,L]$. 
 
In this work, we address the problem of efficient multi-modal adaptation of a transformer-based tumor segmentation model. Suppose that a healthcare institution already has access to a pre-trained segmentation model based on CT scans. If new data containing both CT and PET scans becomes available, adapting the existing model to effectively utilize the PET information is not straightforward. We explore whether recent developments in parameter-efficient fine-tuning (PEFT) techniques such as Low-Rank Adaptation (LoRA) \cite{hu2021lora} can be leveraged to bridge the above gap. LoRA is one of the most PEFTs for adapting large language models like GPT-3 (with $175$ billion parameters) to downstream tasks. It is based on the hypothesis that weight updates to the multi-head self-attention (MHSA) layer of a transformer during adaptation have a low intrinsic rank. Hence, LoRA injects trainable low-rank matrices in parallel to the attention layer while keeping the pre-trained model weights frozen. Based on LoRA, we propose a Parameter Efficient Multimodal Adaptation Approach (PEMMA) that makes best use of the existing pre-trained model for CT and progressively integrates additional modalities (e.g., PET). Our contributions are:  

\begin{itemize}
   \item  We propose a method that offers an efficient incorporation of new modalities into an existing model with minimal cross-modal entanglement.
    \item We demonstrate the feasibility of flexibly and efficiently fine-tuning the adapted model when only one modality is available.
     \item We show how our approach can effectively retain the knowledge learned in the previous steps when trained on new data.
\end{itemize}

\section{Methodology}

\noindent \textbf{Problem Statement}: Suppose that a pre-trained transformer-based tumor segmentation model $\mathcal{F}^{C}_{\Theta}: \mathcal{X}_{C} \rightarrow \mathcal{M}$ for CT scans is available, where $\mathcal{F}^{C}$ denotes the uni-modal (CT) model architecture with parameters $\Theta$, $\mathcal{X}_{C}$ is the input space of CT scans, and $\mathcal{M}$ represents the output segmentation mask space. Our \textit{primary goal} is to upgrade this model such that the adapted model $\widetilde{\mathcal{F}}^{CP}_{\Phi}: \mathcal{X}_{C} \times \mathcal{X}_{P} \rightarrow \mathcal{M}$ can utilize both CT and PET scans to achieve \textbf{better segmentation accuracy}, where $\widetilde{\mathcal{F}}^{CP}$ denotes the multi-modal (CT+PET) model architecture with parameters $\Phi$ and  $\mathcal{X}_{P}$ is the input space of PET scans. We also have two secondary objectives: (i) the model adaptation must be \textbf{parameter-efficient}, i.e., the number of parameters in the multi-modal model (denoted as $|\Phi|$) must be close to that of the uni-modal model (denoted as $|\Theta|$), and (ii) the adaptation must \textbf{minimize cross-modal entanglement}, i.e., it should be possible to subsequently fine-tune the multi-modal model $\widetilde{\mathcal{F}}^{CP}_{\Phi}$ using only one modality (either CT or PET scans), without causing catastrophic forgetting of the other modality.

\begin{figure}[t]
    \centering
    \includegraphics[width=\columnwidth]{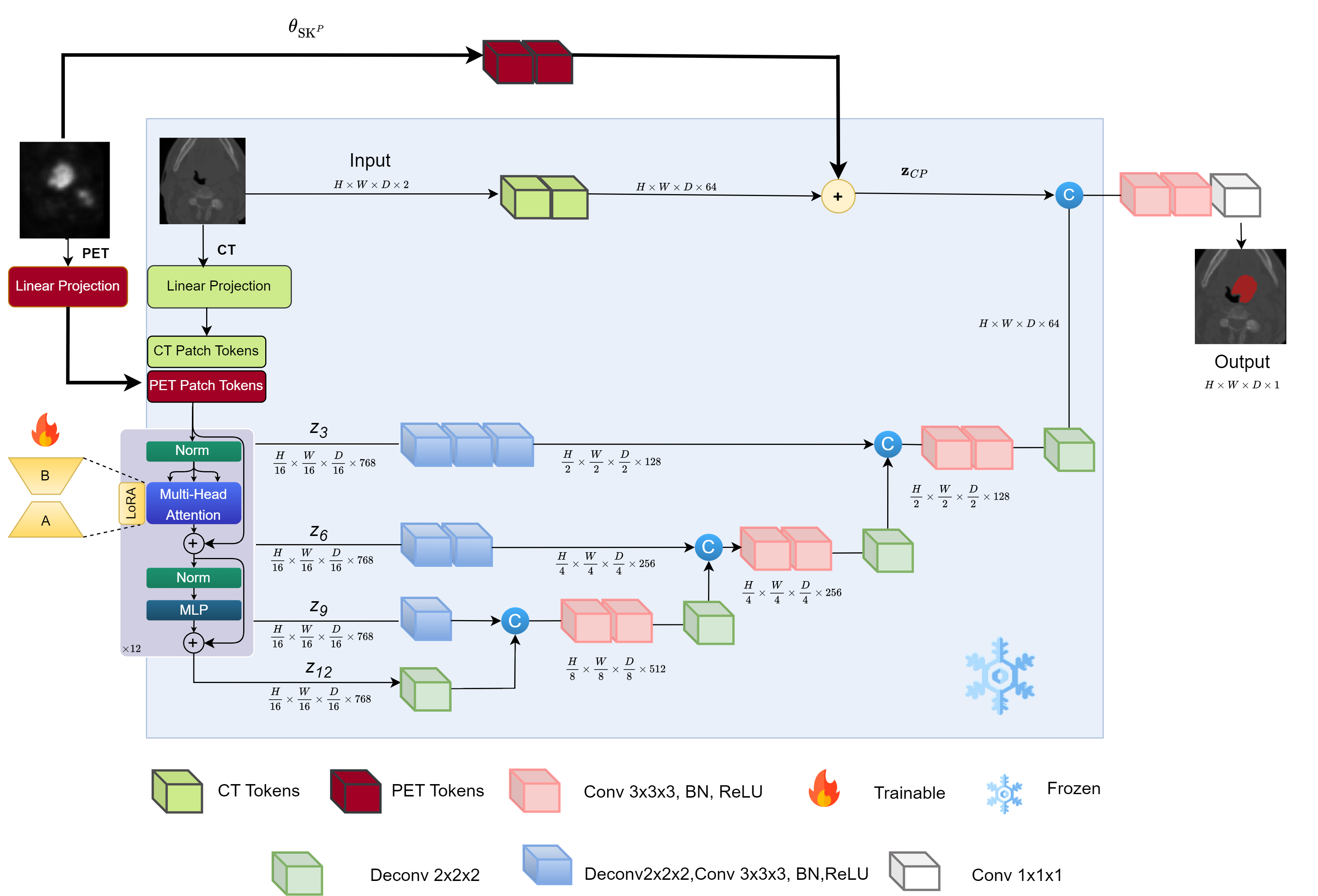}
    \caption{\textit{\textbf{Overview of our proposed architecture  PEMMA:}} At the input level, we separate the path for CT and PET by adding the PET Skip Connection $\theta_{\textrm{SK}}^{P}$. We freeze both the encoder and decoder part of the base UNetr model and introduce LoRA, after each ViT block (x12) as the only trainable layers. 
    }
    \label{fig_main}
\end{figure}

\subsection{Standard Adaptation Methods}

\noindent \textbf{Early Fusion}: The most common approach to train a multi-modal CT+PET segmentation model is to combine the two modalities as different channels of the same input, i.e., create a new multimodal input space $\mathcal{X}_{CP}$, and train a new model  $\widehat{\mathcal{F}}^{CP}_{\Phi}: \mathcal{X}_{CP} \rightarrow \mathcal{M}$ using both the modalities \cite{el-wang2022deep}. When a transformer architecture (e.g., UNETR \cite{hatamizadeh2021unetr}) is used, this involves two main changes to the DNN architecture. Firstly, the uni-modal patch embedding (linear projection) layer that transforms the input data to a set of patch tokens is replaced with a new multi-modal patch embedding layer. Secondly, the direct skip connection between the input and the decoder needs to be modified to accommodate the increase in the number of channels. After making these two architectural changes, the entire model can be learned from the available CT+PET training data.

Specifically, let $\mathbf{x}_{C} \in \mathcal{X}_C$ be the CT image and $\mathcal{T}_0^C = \mathcal{E}_{\theta_{\textrm{PE}}^C}(\mathbf{x}_{C})$ denote the set of CT patch tokens, where $\mathcal{E}_{\theta_{\textrm{PE}}^C}$ is the patch embedding layer of the uni-modal model $\mathcal{F}^{C}_{\Theta}$. Given the PET image $\mathbf{x}_{P} \in \mathcal{X}_P$, we first generate a combined image $\mathbf{x}_{CP} = [\mathbf{x}_{C} || \mathbf{x}_{P}] \in \mathcal{X}_{CP}$, where $||$ denotes channel-wise concatenation. Next, we train a new multi-modal patch embedding layer $\mathcal{E}_{\theta_{\textrm{PE}}^{CP}}$, which generates a new set of CT-PET patch tokens as $\mathcal{T}_0^{CP} = \mathcal{E}_{\theta_{\textrm{PE}}^{CP}}(\mathbf{x}_{CP})$ that are passed to the subsequent transformer blocks in the encoder. Similarly, let $\mathbf{z}_{C} = \mathcal{S}_{\theta_{\textrm{SK}}^{C}}(\mathbf{x}_C)$ be the output of the direct skip connection layer ($\mathcal{S}_{\theta_{\textrm{SK}}^{C}}$) between the input and the decoder in the uni-modal model $\mathcal{F}^{C}_{\Theta}$. Again, we replace $\mathcal{S}_{\theta_{\textrm{SK}}^{C}}$ with a new skip connection layer $\mathcal{S}_{\theta_{\textrm{SK}}^{CP}}$, which outputs $\mathbf{z}_{CP} = \mathcal{S}_{\theta_{\textrm{SK}}^{CP}}(\mathbf{x}_{CP})$.

 The advantage of this adaptation approach is good parameter efficiency because only a minimal number of parameters need to be added to the patch embedding and input skip connection layers to account for the increase in the number of channels. However, the main limitation is that features from both CT and PET modalities get entangled in the combined model. Consequently, any attempt to fine-tune this model further using new data from only one modality (say CT only) will lead to the forgetting of the other modality. Since only minor architectural changes are involved in the adaptation, most of the parameters in the combined model can be initialized using the parameters of the pre-trained uni-modal model. However, for the newly introduced PET-related parameters in the combined model, there are three possible initialization strategies: (i) \textit{random initialization} - initial weights are assigned stochastically, (ii) \textit{zero initialization} - initial weights are set to zero, and (iii) \textit{cross-modal initialization} - the weights of the other modality (CT) in the pre-trained model are re-used. The cross-modal initialization approach can significantly accelerate the learning process and enhance model performance by guiding the PET channel with a pre-learned context of medical imaging, thereby facilitating a more effective CT-PET fusion.

\noindent \textbf{Late Fusion}: Another straightforward approach to incorporate a new modality such as PET is to train a completely new segmentation model $\mathcal{F}^{P}_{\Psi}: \mathcal{X}_{P} \rightarrow \mathcal{M}$ for PET scans,  where $\mathcal{F}^{P}$ denotes the uni-modal (PET) model architecture with parameters $\Psi$, and combine the masks from the two models at the output stage. Let $M_{C} \in \mathcal{M}$ and $M_{P} \in \mathcal{M}$ be the masks produced by the uni-modal CT ($\mathcal{F}^{C}_{\theta}$) and PET ($\mathcal{F}^{P}_{\Psi}$) models, respectively. The combined mask can be computed as:

\begin{equation}
    M_{CP} = w_{C} M_{C} + (1-w_{C}) M_{P},
\end{equation}

\noindent where $w_{C}$ is the weight assigned to the CT modality. While this method offers a great degree of flexibility in dealing with the dynamic availability of the two modalities during training and/or inference, it results in a two-fold increase in the number of parameters ($|\Phi|=|\Theta|+|\Psi|$) and may not provide  optimal accuracy.

\begin{table}[t]
\caption{HECKTOR Dataset description with the division of training and validation data across different centers.}
 \setlength{\tabcolsep}{0.15em} 
{\renewcommand{\arraystretch}{0.7}
\resizebox{\columnwidth}{!}{
\begin{tabular}{@{}ccccc@{}}
\toprule
\textbf{Tasks} & \textbf{Centers} & \textbf{Train Samples } & \textbf{Val Samples} & \textbf{Pet/CT Scanner} \\ \midrule
\multirow{4}{*}{Pre-training} & CHUS & \multirow{4}{*}{203} & \multirow{4}{*}{51} & \\
& CHUV & & & \begin{tabular}[c]{@{}c@{}}GeminiGXL 16 , Philips, Siemens\\ Biograph mCT 40 TOF, \\ Discovery D690 TOF\end{tabular} \\
& CHUM & & & \\
& CHUP & & & \\ \midrule
Multi-Modal Adaptation & MDA & 152 & 44 & Discovery HR, RX, ST, and STE \\ \midrule
CL Task 1 & HGJ & 39 & 15 & Discovery ST, GE Healthcare \\ \midrule
CL Task 2 & HMR & 13 & 4 & Discover ST, GE Healthcare \\ \bottomrule
\label{tab1}
\end{tabular}
}} 
\end{table}






\subsection{Proposed Adaptation Method: PEMMA}



In light of the strengths and weaknesses of existing adaptation methods, we introduce a novel framework (see Fig. \ref{fig_main}) for lightweight adaptation of a uni-modal model into a multi-model modal leveraging the inherent modularity of the transformer architecture. Our proposed framework, referred to as parameter-efficient multi-modal adaptation (PEMMA), is inspired by the concepts of visual prompt tuning (VPT) \cite{vpt} and low-rank adaptation (LoRA) \cite{hu2021lora}. The PEMMA framework has three core components. Firstly, we introduce the new PET modality as a set of visual prompts (or context tokens) to the uni-modal model simply by adding a new patch embedding layer. Secondly, instead of fine-tuning all the parameters of the transformer encoder, we focus only on the attention layers and fine-tune these attention layers through LoRA matrices. Finally, instead of replacing the existing input skip layer in the uni-modal model, we add an additional uni-modal skip layer for PET that operates in parallel to the existing skip layer and the outputs of these two layers are linearly combined. 

Given the PET image $\mathbf{x}_{P}$, we first add a new PET patch embedding layer $\mathcal{E}_{\theta_{\textrm{PE}}^{P}}$, which generates a new set of $N$ PET patch tokens as $\mathcal{T}_0^{P} = \mathcal{E}_{\theta_{\textrm{PE}}^{P}}(\mathbf{x}_{P})$ that are passed to the subsequent transformer blocks in the encoder. Now, the operations of a transformer block can be represented as $\{\mathcal{T}_{\ell}^{C},\mathcal{T}_{\ell}^{P}\} = \mathcal{G}_{\theta_{\ell}}(\{\mathcal{T}_{\ell-1}^{C},\mathcal{T}_{\ell-1}^{P}\})$. Note that the above modification increases the number of tokens processed by the transformer encoder from $N$ to $2N$. However, the decoder in the pre-trained uni-modal model can handle only $N$ tokens. In order to avoid making changes to the decoder architecture, we allow only the $N$ CT tokens from the intermediate transformer blocks ($\mathcal{T}_{\ell}^{C}$) to pass through to the decoder. Refer to Table 2 in the Appendix for Dimensionality Reduction results. It must be emphasized that the self-attention architecture ensures that the knowledge from the PET tokens gets distilled into the CT tokens, even though the PET tokens are ignored by the decoder layers. This is similar to VPT, where the learnable prompts only serve as the context and this contextual information gets distilled into the other tokens, even when the additional prompts are ignored in the end.

The MHSA parameters of a transformer block $\ell$ can be considered as a collection of four weight matrices denoted as $\{\mathbf{W}_{O,\ell},\mathbf{W}_{Q,\ell},\mathbf{W}_{K,\ell},\mathbf{W}_{V,\ell}\}$. In LoRA, the updates to $\mathbf{W}_{Q,\ell}$ and $\mathbf{W}_{V,\ell}$ are decomposed into a pair of low rank matrices $\mathbf{A} \in \mathbb{R}^{r \times d}$ and $\mathbf{B} \in \mathbb{R}^{d \times r}$, where $r$ represents the rank of the two matrices. Let $h_{*,\ell}$ and $\tilde{h}_{*,\ell}$ be the input and output, respectively, of an attention layer in the $\ell^{\text{th}}$ block. Then, LoRA operation can be summarized as: 

\begin{equation}
    \label{eq:processing_lora}
    \begin{split}
        \tilde{h}_{Q, \ell} &= \mathbf{W}_{Q,\ell}  h_{Q, \ell} + \alpha \mathbf{B}_{Q,\ell}  \mathbf{A}_{Q,\ell}   h_{Q, \ell} \\ 
         \tilde{h}_{V, \ell} &= \mathbf{W}_{V,\ell}  h_{V, \ell} + \alpha  \mathbf{B}_{V,\ell}  \mathbf{A}_{V,\ell}   h_{V, \ell}
    \end{split}
\end{equation} 

\noindent where $\alpha$ is a fixed scalar. Note that parameter-efficiency is achieved by allowing $\theta_{\textrm{LoRA}} = \{\mathbf{A}_{Q,\ell},\mathbf{A}_{V,\ell},\mathbf{B}_{Q,\ell},\mathbf{B}_{V,\ell}\}_{\ell=1}^L$ as the only learnable parameters and freezing the rest of the parameters in the transformer encoder. 

Finally, in contrast to the early fusion approach where $\mathcal{S}_{\theta_{\textrm{SK}}^{C}}$ is replaced with $\mathcal{S}_{\theta_{\textrm{SK}}^{CP}}$, we leave  $\mathcal{S}_{\theta_{\textrm{SK}}^{C}}$ untouched and introduce an additional parallel path $\mathcal{S}_{\theta_{\textrm{SK}}^{P}}$ for the PET image. Let $\mathbf{z}_{P} = \mathcal{S}_{\theta_{\textrm{SK}}^{P}}(\mathbf{x}_P)$ be the output of the additional direct skip connection layer introduced for the PET modality. In this case, the combined output of the input skip layer is $\mathbf{z}_{CP} = \mathbf{z}_{C} + \beta \mathbf{z}_{P}$, where $\beta$ is the weight assigned to the PET modality. The main motivation for the introduction of new patch embedding ( $\mathcal{E}_{\theta_{\textrm{PE}}^{P}}$) and input skip layers ($\mathcal{S}_{\theta_{\textrm{SK}}^{P}}$) for the PET modality (rather than replacing them as in early fusion) is to minimize cross-modal entanglement. With the introduction of these additional layers, it is possible to subsequently fine-tune the multi-modal model using only one of the modalities, without affecting the model's ability to handle the other modality. 

\noindent \textbf{Flexible Training and Inference Strategy}: The PEMMA framework introduces three new parameters $\theta_{\textrm{PE}}^{P}$, $\theta_{\textrm{LoRA}}$, and $\theta_{\textrm{SK}}^{P}$. When adapting the uni-modal model to the multimodal scenario, both CT and PET training data is required and all the new parameters $\{\theta_{\textrm{PE}}^{P},\theta_{\textrm{LoRA}},\theta_{\textrm{SK}}^{P}\}$ are learned, while the parameters of the pre-trained unimodal model $\Theta$ are completely frozen. Thus, the parameters of the multi-modal model are $\Phi = \{\Theta,\theta_{\textrm{PE}}^{P},\theta_{\textrm{LoRA}},\theta_{\textrm{SK}}^{P}\}$ and $|\Phi|$ is only marginally higher than $|\Theta|$ (hence, parameter-efficient). Subsequently, if new data is available to update the multi-modal model, only $\theta_{\textrm{LoRA}}$ needs to be updated and all other parameters can be frozen. This allows the flexibility of fine-tuning the multi-modal model using one or both modalities (if a modality is unavailable, the corresponding input can be set to zero). Similarly, the multi-modal allows flexible inference - it can effectively utilize both modalities when they are available, but can also be applied to only a single modality (albeit with some degradation in the segmentation accuracy).


\begin{table}[!ht]
 \centering
 \caption{Results for the different Adaptation Approaches performances on the Training and Inference Modalities. We compare the number of trainable parameters with the UNETR model where $\Phi$= 92.58M params. We compute the Tumor, Lymph, and Average Dice scores for each experiment. C=CT;P=PET;CP=CT+PET.}
 \resizebox{\columnwidth}{!}{%
   \setlength{\tabcolsep}{0.55em} 
   {\renewcommand{\arraystretch}{1.5}
     \begin{tabular}{@{}ccccc|cccccc|cccccc@{}} 
       \toprule
       \rowcolor[HTML]{C0C0C0} 
       \textbf{Dataset} & \multicolumn{1}{c}{\cellcolor[HTML]{C0C0C0}$\longrightarrow$} & \multicolumn{3}{c|}{\cellcolor[HTML]{C0C0C0}\textbf{Multimodal Adaptation}} & \multicolumn{6}{|c|}{\cellcolor[HTML]{C0C0C0}\textbf{New Dataset 1}} & \multicolumn{6}{c}{\cellcolor[HTML]{C0C0C0}\textbf{New Dataset 2}} \\
       \textbf{Train Modalities} & \multicolumn{1}{c}{$\longrightarrow$} & \multicolumn{3}{c|}{\textbf{CP}} & \multicolumn{3}{c|}{\textbf{C}} & \multicolumn{3}{c|}{\textbf{CP}} & \multicolumn{3}{c|}{\textbf{C}} & \multicolumn{3}{c}{\textbf{CP}} \\
       \rowcolor[HTML]{C0C0C0} 
       \textbf{Infer Modalities} & \multicolumn{1}{c}{\cellcolor[HTML]{C0C0C0}$\longrightarrow$} & \multicolumn{1}{c}{\cellcolor[HTML]{C0C0C0}\textbf{CP}} & \multicolumn{1}{c}{\cellcolor[HTML]{C0C0C0}\textbf{C}} & \multicolumn{1}{c|}{\cellcolor[HTML]{C0C0C0}\textbf{P}} & \multicolumn{1}{c}{\cellcolor[HTML]{C0C0C0}\textbf{CP}} & \multicolumn{1}{c}{\cellcolor[HTML]{C0C0C0}\textbf{C}} & \multicolumn{1}{c|}{\cellcolor[HTML]{C0C0C0}\textbf{P}} & \multicolumn{1}{c}{\cellcolor[HTML]{C0C0C0}\textbf{CP}} & \multicolumn{1}{c}{\cellcolor[HTML]{C0C0C0}\textbf{C}} & \multicolumn{1}{c|}{\cellcolor[HTML]{C0C0C0}\textbf{P}} & \multicolumn{1}{c}{\cellcolor[HTML]{C0C0C0}\textbf{CP}} & \multicolumn{1}{c}{\cellcolor[HTML]{C0C0C0}\textbf{C}} & \multicolumn{1}{c|}{\cellcolor[HTML]{C0C0C0}\textbf{P}} & \multicolumn{1}{l}{\cellcolor[HTML]{C0C0C0}\textbf{CP}} & \multicolumn{1}{c}{\cellcolor[HTML]{C0C0C0}\textbf{C}} & \multicolumn{1}{c}{\cellcolor[HTML]{C0C0C0}\textbf{P}} \\ 
       \midrule
       & \large Tumor & \large0.69 & \large0.37 & \large0.43 & \large0.67 & \large0.45 & \multicolumn{1}{c|}{\large0.32} & \large0.40 & \large0.36 & \large0.23 & \large0.68 & \large0.43 & \multicolumn{1}{c|}{\large0.05} & \large0.47 & \large0.46 & \large0.24 \\
       & \large Lymph & \large0.64 & \large0.51 & \large0.51 & \large0.61 & \large0.56 & \multicolumn{1}{c|}{\large0.21} & \large0.48 & \large0.46 & \large0.27 & \large0.56 & \large0.52 & \multicolumn{1}{c|}{\large0.32} & \large0.30 & \large0.34 & \large0.11 \\
       \multirow{-3}{*}{\begin{tabular}[c]{@{}c@{}}\textbf{Late Fusion}\\ \textcolor{red}{(\textbf{\large params=2} $\Phi$)}\end{tabular}} & \large Avg & \large0.65 & \large0.43 & \large0.47 & \large0.64 & \large0.49 & \multicolumn{1}{c|}{\large0.17} & \large0.44 & \large0.39 & \large0.25 & \large0.62 & \textbf{\large0.48} & \multicolumn{1}{c|}{\large0.19} & \large0.39 & \large0.35 & \large0.18 \\
       \midrule
       & \large Tumor & \large0.65 & \large0.41 & \large0.68 & \large0.76 & \large0.71 & \multicolumn{1}{c|}{\large0.17} & \large0.81 & \large0.42 & \large0.30 & \large0.82 & \large0.34 & \multicolumn{1}{c|}{\large0.01} & \large0.63 & \large0.35 & \large0.42 \\
       & \large Lymph & \large0.63 & \large0.48 & \large0.64 & \large0.63 & \large0.64 & \multicolumn{1}{c|}{\large0.01} & \large0.56 & \large0.48 & \large0.37 & \large0.58 & \large0.38 & \multicolumn{1}{c|}{\large0.04} & \large0.56 & \large0.35 & \large0.25 \\
       \multirow{-3}{*}{\begin{tabular}[c]{@{}c@{}} \textbf{Early Fusion}\\  \textcolor{red}{(\textbf{\large params=1.0043} $\Phi$)}\end{tabular}} & \large Avg & \textbf{\large0.67} & \large0.43 & \large0.65 & \large0.70 & \large0.68 & \multicolumn{1}{c|}{\large0.09} & \large0.69 & \large0.43 & \large0.31 & \large0.07 & \large0.36 & \multicolumn{1}{c|}{\large0.03} & \large0.60 & \large0.35 & \textbf{\large0.34} \\
       \midrule
       & \large Tumor & \large0.67 & \large0.41 & \large0.64 & \large0.82 & \large0.70 & \multicolumn{1}{c|}{\large0.29} & \large0.81 & \large0.43 & \large0.31 & \large0.86 & \large0.38 & \multicolumn{1}{c|}{\large0.34} & \large0.64 & \large0.35 & \large0.39 \\
       & \large Lymph & \large0.60 & \large0.50 & \large0.63 & \large0.61 & \large0.72 & \multicolumn{1}{c|}{\large0.30} & \large0.57 & \large0.48 & \large0.40 & \large0.68 & \large0.50 & \multicolumn{1}{c|}{\large0.26} & \large0.57 & \large0.48 & \large0.26 \\
       \begin{tabular}[c]{@{}c@{}} \textbf{PEMMA (Ours)}\\ \textcolor{red}{(\textbf{\large params=0.08} $\Phi$)} \end{tabular} & \large Avg & \large0.63 & \textbf{\large0.44} & \textbf{\large 0.65} & \textbf{\large0.74} & \textbf{\large0.69} & \multicolumn{1}{c|}{\textbf{\large0.28}} & \textbf{\large0.72} & \textbf{\large0.45} & \textbf{\large0.33} & \textbf{\large0.75} & \large0.43 & \multicolumn{1}{c|}{\textbf{\large0.31}} & \textbf{\large0.63} & \textbf{\large0.41} & \large0.32 
       \\
       \bottomrule
     \end{tabular}
   }
 }
 \label{tab2}
\end{table}

\section{Experimental Set-up}

\subsection{Dataset Description}



\noindent\textbf{Pre-processing:} The dataset \cite{hecktor-andrearczyk2022overview} is publicly available on the MICCAI 2022 HEad and neCK TumOR (HECKTOR) challenge website. In total, the dataset compromises 522 samples, with a breakdown provided in Table 1 \ref{tab1} detailing distribution across various centers and specifying the scanner types utilized for scan acquisition.
The training enhancements applied to both scans involve extracting four random crops sized 96×96×96. These augmentations aim to diversify and represent the training data more comprehensively For further details on the pre-processing steps refer to Table 1 in the Appendix.  

\noindent\textbf{Implementation Details:}
Our approach runs on PyTorch version 2.1.0 utilizing the MONAI Library \cite{cardoso2022monai}. We train all models for a maximum of 18k steps and select the best model based on the highest average dice score on validation set. We use a learning rate of 1e-4  and weight decay of 1e-5 for all experiments and train using AdamW optimizer with a batch size of 2. Our entire pipeline runs on a single Nvidia A6000 RTX 48GB GPU.

\hfill

\section{Results and Discussion}

Our experimental approach leverages  four centers of the HECKTOR Dataset (CHUM,CHUV,CHUP and CHUS) to construct our initial pre-trained unimodal (CT only), as shown in Table 1 \ref{tab1}. We then fine-tune both modalities in the multi-modal adaptation setting, leveraging data from MDA center.

We test our method on both joint (CT+PET) and separate modalities. We compare our findings with the standard adaptation approaches defined previously: Early and Late. We observe that PEMMA yields results on par with the early fusion approach,  yet it is \textbf{12x}  more efficient. To adapt to new datasets, we introduce the CT and PET scans from two new centers (HGJ and HMR) and design two training scenarios that reflect real-world medical environments: one model trained on CT only (when PET data is missing), and another with both modalities available. PEMMA surpasses both early and late fusion techniques across the various adaptation and inference strategies, particularly in enhancing the extra-modality (PET) tuning, evidenced by a remarkable increase of \textbf{+19\%} and \textbf{+28\%} in Average Dice scores for new datasets 1 and 2, respectively. 

These outcomes highlight LoRA's effectiveness in progressively integrating additional modality information while significantly reducing the need for comprehensive model retraining, cutting down trainable parameters by \textbf{92\%}.
Finally, the proposed approach can also be useful in a continual learning
setting, where new tasks with different data distributions may be introduced
sequentially. Since LoRA is highly parameter-efficient, only very few parameters( \(|\theta_{\text{LoRA}}| \ll |\Phi|\) )
need to be updated for each task. 
Hence, it becomes much more feasible to store all task-specific parameters in memory and achieve good forward transfer as well as non-negative backward transfer (thereby circumventing
catastrophic forgetting). We, therefore, successfully achieve the parameter efficiency and minimization of the cross-modality entanglement goals.

\newcommand{\inc}[1]{\ensuremath{_{\text{\textcolor{green}{(+#1)}}}}}
\newcommand{\dec}[1]{\ensuremath{_{\ text{\textcolor{magenta}{(-#1)}}}}}
\newcommand{\cmark}{\ding{51}}%
\newcommand{\xmark}{\ding{55}} %
 
\definecolor{Gray}{gray}{0.90}
\definecolor{white}{rgb}{1.0, 1.0, 1.0}
 
\definecolor{Lightgreen}{RGB}{218, 246, 230 }
\newcolumntype {a}{>{\columncolor{Lightgreen}}c}
\definecolor{Gray}{gray}{0.90}
\newcommand{\theHalgorithm}{\arabic{algorithm}}
\definecolor{label1}{rgb}{0.76,0.59,0.77}
\definecolor{label2}{rgb}{0.28,0.5,0.72}
\definecolor{label3 }{rgb}{0.33,0.63,0.36}
\definecolor{label4}{rgb}{0.79,0.4,0.17}
\definecolor{label5}{rgb}{0.94,0.53,0.2}
\definecolor{label6}{rgb}{ 0.72,0.86,0.59}
\definecolor{label7}{rgb}{1,1,0.65}
\definecolor{label8}{rgb}{0.93,0.62,0.61}
\definecolor{label9}{rgb}{0.4,0.15,0.33 }
\definecolor{label10}{rgb}{0.75,0.21,0.29}
\definecolor{label11}{rgb}{0.35,0.73,0.8}
\definecolor{label12}{rgb}{0.94,0.9,0.32}
\definecolor{label13 }{rgb}{0.96,0.76,0.48}
 
\newsavebox{\spleen}
\savebox{\spleen}{\textcolor{label1}{\rule{1.5in}{1.5in}}}
 
\newsavebox{\rkid}
\savebox{\rkid}{\textcolor{label2}{\rule{1.5in}{1.5in}}}
 
\newsavebox{\lkid}
 
\newsavebox{\gall}
\savebox{\gall}{\textcolor{label4}{\rule{1.5in}{1.5in}}}
 
\newsavebox{\eso}
\savebox{\eso}{\textcolor{label5}{\rule{1.5in}{1.5in}}}
 
\newsavebox{\liver}
\savebox{\liver}{\textcolor{label6}{\rule{1.5in}{1.5in}}}
 
\newsavebox{\sto}
\savebox{\sto}{\textcolor{label7}{\rule{1.5in}{1.5in}}}
 
\newsavebox{\aorta}
\savebox{\aorta}{\textcolor{label8}{\rule{1.5in}{1.5in}}}
 
\newsavebox{\ivc}
\savebox{\ivc}{\textcolor{label9}{\rule{1.5in}{1.5in}}}
 
\newsavebox{\veins}
\savebox{\veins}{\textcolor{label10}{\rule{1.5in}{1.5in}}}
 
\newsavebox{\panc}
\savebox{\panc}{\textcolor{label11}{\rule{1.5in}{1.5in}}}
 
\newsavebox{\rad}
\savebox{\rad}{\textcolor{label12}{\rule{1.5in}{1.5in}}}
 
\newsavebox{\lad}
 
\newsavebox{\RV}
\savebox{\RV}{\textcolor{blue}{\rule{1.5in}{1.5in}}}
\newcommand{\muz}[1]{{\textcolor{ network}{#1}}}


%
\newcolumntype{x}{>{\scriptsize\raggedright\hspace{0pt}}X}
\makeatother

\section{Conclusion}
In this study, we introduce a Parameter-Efficient Multimodality Adaptation method to enhance the proposed model's adaptability and efficiency across various data sources and modalities without substantial retraining or a major increase in the number of the model's parameters. This technique allows our model to retain past knowledge and further learn from new data modalities, improving its ability to process evolving multimodal information. Looking ahead, we aim to test this method with other medical imaging types, like MRI, and expand our research to include diverse datasets. Further evaluation is needed to assess the model's reliability and flexibility in a wider range of medical imaging scenarios, potentially creating more effective diagnostic tools. Finally, an interesting research avenue that could be built on our multimodality adaptation approach is to investigate Parameter-Efficient Multi-disease Adaptation to allow DNNs to adapt efficiently and effectively when trained on data from different diseases.

\bibliographystyle{splncs04}
\bibliography{references}

\begin{thebibliography}{10}
\providecommand{\url}[1]{\texttt{#1}}
\providecommand{\urlprefix}{URL }
\providecommand{\doi}[1]{https://doi.org/#1}

\bibitem{hecktor-andrearczyk2022overview}
Andrearczyk, V., Oreiller, V., Boughdad, S., Rest, C.C.L., Elhalawani, H., Jreige, M., Prior, J.O., Vallières, M., Visvikis, D., Hatt, M., Depeursinge, A.: Overview of the hecktor challenge at miccai 2021: Automatic head and neck tumor segmentation and outcome prediction in pet/ct images (2022)

\bibitem{cardoso2022monai}
Cardoso, M.J., Li, W., Brown, R., Ma, N., Kerfoot, E., Wang, Y., Murrey, B., Myronenko, A., Zhao, C., Yang, D., Nath, V., He, Y., Xu, Z., Hatamizadeh, A., Myronenko, A., Zhu, W., Liu, Y., Zheng, M., Tang, Y., Yang, I., Zephyr, M., Hashemian, B., Alle, S., Darestani, M.Z., Budd, C., Modat, M., Vercauteren, T., Wang, G., Li, Y., Hu, Y., Fu, Y., Gorman, B., Johnson, H., Genereaux, B., Erdal, B.S., Gupta, V., Diaz-Pinto, A., Dourson, A., Maier-Hein, L., Jaeger, P.F., Baumgartner, M., Kalpathy-Cramer, J., Flores, M., Kirby, J., Cooper, L.A.D., Roth, H.R., Xu, D., Bericat, D., Floca, R., Zhou, S.K., Shuaib, H., Farahani, K., Maier-Hein, K.H., Aylward, S., Dogra, P., Ourselin, S., Feng, A.: Monai: An open-source framework for deep learning in healthcare (2022)

\bibitem{early-fusion10.1055/a-1542-6211}
Farag, S., IJzerman, N.S., Houdijk, M.P., Reyners, A., Arens, A., Grünhagen, D.J., Desar, I.M., Gelderblom, H., Steeghs, N., Geus-Oei, L.d.: Early response evaluation using 18f-fdg-pet/ct does not influence management of patients with metastatic gastrointestinal stromal tumors (gist) treated with palliative intent. Nuklearmedizin - NuclearMedicine  \textbf{60},  411--416 (2021). \doi{10.1055/a-1542-6211}

\bibitem{hatamizadeh2021unetr}
Hatamizadeh, A., Tang, Y., Nath, V., Yang, D., Myronenko, A., Landman, B., Roth, H., Xu, D.: Unetr: Transformers for 3d medical image segmentation (2021)

\bibitem{hu2021lora}
Hu, E.J., Shen, Y., Wallis, P., Allen-Zhu, Z., Li, Y., Wang, S., Wang, L., Chen, W.: Lora: Low-rank adaptation of large language models (2021)

\bibitem{fusion-huang2024vision}
Huang, H., Qiu, L., Yang, S., Li, L., Nan, J., Li, Y., Han, C., Zhu, F., Zhao, C., Zhou, W.: Vision transformer-based multimodal feature fusion network for lymphoma segmentation on pet/ct images (2024)

\bibitem{vpt}
Jia, M., Tang, L., Chen, B.C., Cardie, C., Belongie, S., Hariharan, B., Lim, S.N.: Visual prompt tuning (2022)

\bibitem{doi:fusion-10.1080/0284186X.2021.1949034}
Jintao~Ren, Jesper Grau~Eriksen, J.N., Korreman, S.S.: Comparing different ct, pet and mri multi-modality image combinations for deep learning-based head and neck tumor segmentation. Acta Oncologica  \textbf{60}(11),  1399--1406 (2021). \doi{10.1080/0284186X.2021.1949034}, \url{https://doi.org/10.1080/0284186X.2021.1949034}, pMID: 34264157

\bibitem{late-fusion10.1117/12.878067}
Kadoury, S., Wood, B.J., Venkatesan, A.M., Dalal, S., Xu, S., Kruecker, J.: Accuracy assessment of an automatic image-based pet/ct registration for ultrasound-guided biopsies and ablations. SPIE Proceedings  (2011). \doi{10.1117/12.878067}

\bibitem{gnn-PENG2024106137}
Peng, J., Peng, L., Zhou, Z., Han, X., Xu, H., Lu, L., Lv, W.: Multi-level fusion graph neural network: Application to pet and ct imaging for risk stratification of head and neck cancer. Biomedical Signal Processing and Control  \textbf{92},  106137 (2024). \doi{https://doi.org/10.1016/j.bspc.2024.106137}, \url{https://www.sciencedirect.com/science/article/pii/S1746809424001952}

\bibitem{TMSS-10.1007/978-3-031-16449-1_31}
Saeed, N., Sobirov, I., Al~Majzoub, R., Yaqub, M.: Tmss: An end-to-end transformer-based multimodal network for segmentation and survival prediction. In: Wang, L., Dou, Q., Fletcher, P.T., Speidel, S., Li, S. (eds.) Medical Image Computing and Computer Assisted Intervention -- MICCAI 2022. pp. 319--329. Springer Nature Switzerland, Cham (2022)

\bibitem{el-wang2022deep}
Wang, Y.e.a.: Deep learning based time-to-event analysis with pet, ct and joint pet/ct for head and neck cancer prognosis. Computer Methods and Programs in Biomedicine  \textbf{222},  106948 (2022). \doi{10.1016/j.cmpb.2022.106948}

\end{thebibliography}







\end{document}


\title{PEMMA: Parameter-Efficient Multi-Modal Adaptation for Medical Image Segmentation}

%
\titlerunning{}
%
\author{*}
%

\section{Dataset Preprocessing}

\begin{table}[htbp]
\caption{Description of the pre-processing and Augmentation details performed on the HECKTOR Dataset. }
\resizebox{0.7\columnwidth}{!}{%
\begin{tabular}{@{}cccc@{}}
\rowcolor[HTML]{CBCEFB} 
{\ul \textbf{Augmentations}}                  & {\ul \textbf{Axis}}        & {\ul \textbf{Probability}} & {\ul \textbf{Size}}           \\ \midrule
\multicolumn{1}{|c|}{Orientation}             & \multicolumn{1}{c|}{PLS}   & \multicolumn{1}{c|}{-}     & \multicolumn{1}{c|}{-}        \\ \midrule
\multicolumn{1}{|c|}{CT/PET Concatenation}    & \multicolumn{1}{c|}{1}     & \multicolumn{1}{c|}{-}     & \multicolumn{1}{c|}{-}        \\ \midrule
\multicolumn{1}{|c|}{Random Crop}             & \multicolumn{1}{c|}{-}     & \multicolumn{1}{c|}{0.5}   & \multicolumn{1}{c|}{96x96x96} \\ \midrule
\multicolumn{1}{|c|}{Random Flip}             & \multicolumn{1}{c|}{x,y,z} & \multicolumn{1}{c|}{0.2}   & \multicolumn{1}{c|}{-}        \\ \midrule
\multicolumn{1}{|c|}{Rotate by 90 (up to 3x)} & \multicolumn{1}{c|}{x,y}   & \multicolumn{1}{c|}{0.2}   & \multicolumn{1}{c|}{-}        \\ \bottomrule
\end{tabular}%
}
\label{Table A}

\end{table}

\section{Dimensionality reduction}

\begin{table}[!ht]
\caption{we explore different dimensionality reduction techniques while passing 2NxD in the encoder. We are faced by the need to keep the dimensions of the input of the decoder part as Nxd, hence we conduct one experiment where take alternative tokens from both modalities,one where we take tokens coming only from CT, and another one where where we only pass the CT tokens to the decoder. The latter one shows better dice scores, consequently we adopt this dimeninality reduction for all the subsequent experiments.}
\resizebox{0.6\columnwidth}{!}{%
\begin{tabular}{@{}cccc@{}}
\toprule
 MDA                       & Mix        &  CT only           & PET only  \\ \midrule
\multicolumn{1}{c|}{Tumor Dice} & \multicolumn{1}{c|}{0.47} & \multicolumn{1}{c|}{0.63} & 0.30             \\
\multicolumn{1}{c|}{Lymph Dice} & \multicolumn{1}{c|}{0.20} & \multicolumn{1}{c|}{0.57} & 0.06             \\
\multicolumn{1}{c|}{Avg Dice}   & \multicolumn{1}{c|}{0.49} & \multicolumn{1}{c|}{0.62} & 0.27             \\ \bottomrule
\end{tabular}%
}
\label{Table B}

\end{table}